\documentclass{article}
\usepackage{hyperref}
\usepackage[left=1in,right=1in,top=1in,bottom=1in]{geometry}
\usepackage{natbib}
\usepackage{graphicx}
\usepackage{epstopdf}

\begin{document}
\thispagestyle{empty}
\begin{center}
\vspace*{0cm} \large\textbf{Towards building a first northern-sky sample of \lq Extremely Inverted Spectrum Extragalactic Radio Sources (EISERS) \rq}	\\
\vspace*{0.8cm} \small{{Mukul Mhaskey$^{1}$, Gopal-Krishna$^{2}$ and Surajit Paul$^{1}$}}\\
 
\vspace*{0.4cm} \it \tiny{$^{1}$Department of Physics, Savitribai Phule Pune Unversity, Ganeshkhind, Pune 411007, India\\
 $^{2}$Aryabhatta Research Institute of Observational Sciences (ARIES), Manora Peak, Nainital $-$ 263129, India}\\

\end{center}

\begin{abstract}
We present here an extension of our search for EISERS (Extremely Inverted Spectrum Extragalactic Radio Sources) to the northern hemisphere. With an inverted radio spectrum of slope $\alpha$ $>$ + 2.5, these rare sources would either require a non-standard particle acceleration mechanism (in the framework of synchrotron self-absorption hypothesis), or a severe free-free absorption which attenuates practically all of their synchrotron radiation at metre wavelengths. A list of 15 EISERS candidates is presented here. It was assembled  by applying a sequence of selection filters, starting with the two available large-sky radio surveys, namely the WENSS (325 MHz) and the ADR-TGSS (150 MHz). These surveys offer the twin advantages of being fairly deep (typical rms $<$ 10 mJy/beam) and having a sub-arcminute resolution. Their zone of overlap spreads over 1.3$\pi$ steradian in the northern hemisphere.  Radio spectra are presented for the entire sample of 15 EISERS candidates, of which 8 spectra are of GPS type. Eleven members of the sample are associated with previously known quasars. Information on the parsec-scale radio structure, available for several of these sources, is also summarized.  
\end{abstract}

\section{Introduction}

Sensitive large-area radio surveys, such as the FIRST \citep{Helfand2015} and NVSS \citep{Condon1998}, combined with their optical counterpart, SDSS \citep{Alam2015} have enabled the building of large samples for detailed studies of extragalactic radio sources e.g. \citep{Best2005, Coziol2017}. In recent years, such samples have been increasingly used not only for finding, rare types of radio galaxies, such as X-shaped radio sources, e.g. (\citet{Cheung2013, Roberts2018, GK2012} and references therein), giant radio galaxies e.g. \citep{Dabhade2017}, recurrently active radio galaxies e.g. \citep{Kuzmicz2017} and HYMORS \citep{Gopal-Krishna2000, Kapinska2017}, but also species like FR0 radio galaxies \citep{Baldi2015} and `Compact Symmtric Objects' (e.g., \citet{Wilkinson1994, Conway1994, Gugliucci2005, Phillips1982}) whose abundance had remained scarcely recognised. A major advance in recent times is the availability of fairly deep large-area surveys at metre wavelengths, such as the ADR-TGSS \citep{Intema2017} and the GLEAM survey \citep{Wayth2015, Hurley-Walker2017}. 
A head-start in this direction had already been made with the Westerbork surveys, WENSS at 325 MHz \citep{deBruyn2000} and WISH at 352 MHz \citep{DeBreuck2002}, as well as the VLA 74 MHz survey \citep{Lane2014}. 

A few years ago, we combined the TIFR.GMRT.SKY.SURVEY (TGSS/DR5) at 150 MHz with the 352 MHz WISH survey, to search for radio galaxies whose radio spectrum shows an ultra-sharp turnover, such that the inverted spectrum has a slope greater than the critical value $\alpha_c$ $=$ $+$2.5, the theoretical limit for synchrotron self-absorption (SSA)\citep{GopalKrishna2014} (henceforth Paper I). This limit cannot be violated even by a perfectly homogeneous source which emits incoherent synchrotron radiation from relativistic electrons whose energy distribution has the canonical (i.e., power-law) shape \citep{Slish1963,Scheuer1968,Rybicki1979}. Consequently, such sources would either be having a non-standard electron energy distribution \citep{Rees1967}, or suffer a severe free-free absorption (FFA) at meter wavelengths, as discussed in Paper I where they were termed `Extremely Inverted Spectrum Extragalactic Radio Sources' (EISERS). The 7 EISERS candidates reported in Paper I were subsequently followed up with the Giant Metrewave Radio Telescope - GMRT \citep{Swarup1991}, by making quasi-simultaneous observations at 150, 325, 610 and 1400 MHz (Mhaskey et al. 2018, submitted, henceforth Paper II). This and a similarly targeted independent program, based on the GLEAM survey in which flux densities are measured simultaneously at several narrow bands in the 72-231 MHz range \citep{Callingham2017}, have together been able to find just a couple of EISERS, which underscores the extreme rarity of such radio sources. The aim of the present study is to extend 
the search for EISERS to the northern hemisphere, by combining the available two large-area, high-resolution radio surveys, namely the recently published ADR-TGSS at 150 MHz \citep{Intema2017} and the pre-existing WENSS at 325 MHz \citep{deBruyn2000}. The pairing of these large-area radio surveys was preferred in view of their fairly high sensitivity (typical rms $<$ 10 mJy/beam) and a sub-arcminute resolution.  

\section{The selection of EISERS candidates in the northern sky}

The overlap region between the two basic radio surveys (sect. 1), viz., the ADR-TGSS (150 MHz) and WENSS (325 MHz), spans nearly 1.3$\pi$ steradian on the sky (1/3rd of the total sky) and contains 229420 radio sources in the WENSS catalogue which covers the declination range north of +28$^\circ$. Out of this large sample, we first extracted a subset containing all those 35064 sources which belong to the morphological type \lq S\rq~(i.e., single) and are stronger than 150 mJy at 325 MHz, as per the WENSS catalogue. For these shortlisted WENSS sources, we then looked for their ADR-TGSS counterpart, taking a search radius of 20 arcsec. A total of 33707 WENSS sources passed this filter and for them the separation between the WENSS and ADR-TGSS positions is found to have a median value of 2.10$\pm$0.02 arcsec, which is consistent with the quoted positional uncertainties for the relatively weak sources in the two surveys (rms $\sim$ 1 arcsec for WENSS and $\sim$ 2 arcsec for ADR-TGSS). Next, for the WENSS sources without
a counterpart in the ADR-TGSS, we tentatively assumed an upper limit of 30 mJy at 150 MHz, which is nearly 6 times the typical rms uncertainty at such flux levels. Thereafter, for each shortlisted WENSS source we computed the spectral index $\alpha$ (150-325 MHz) (or, lower limit to $\alpha$, in the cases of non-detection at 150 MHz). The availability of this information for each source allowed us to select those having $\alpha$(150-325 MHz) $>$ +2.0. A total of 113 sources were thus shortlisted.
     
In the next round, we inspected radio images of each of the 113 EISERS candidates in the ADR-TGSS, WENSS and NVSS surveys. All cases showing extended emission were flagged and deleted (16 sources). Sources lying in the galactic plane or known to be H-II regions were also deleted (25 sources). In order to refine the estimated values of $\alpha$ (150-325 MHz) for the sources lacking a counterpart in the ADR-TGSS catalogue at 150 MHz (for which we had assumed a fixed upper limit of 30 mJy, see above), we then measured the rms noise on the ADR-TGSS images of individual sources and five times the rms noise was set as the new upper limit for the 150 MHz flux density. This new limit was used to recompute the lower limit to $\alpha$ (150-325 MHz) for the sources still undetected in the 150 MHz ADR-TGSS images. We note that many sources which are not listed in the ADR-TGSS catalogue could in fact be seen in the downloaded images (using the CASA software). Consequently, it became possible to calculate for them the actual value of $\alpha$ (150-325 MHz) and the lower limit estimated earlier was hence discarded. We note that for 10 out of the 46 shortlisted sources, 150 MHz counterparts could thus be found, thereby shrinking considerably the list of sources with just a lower limit to $\alpha$ (150-325 MHz).

\begin{figure}[!htbp]
\centering
\includegraphics[scale=.45]{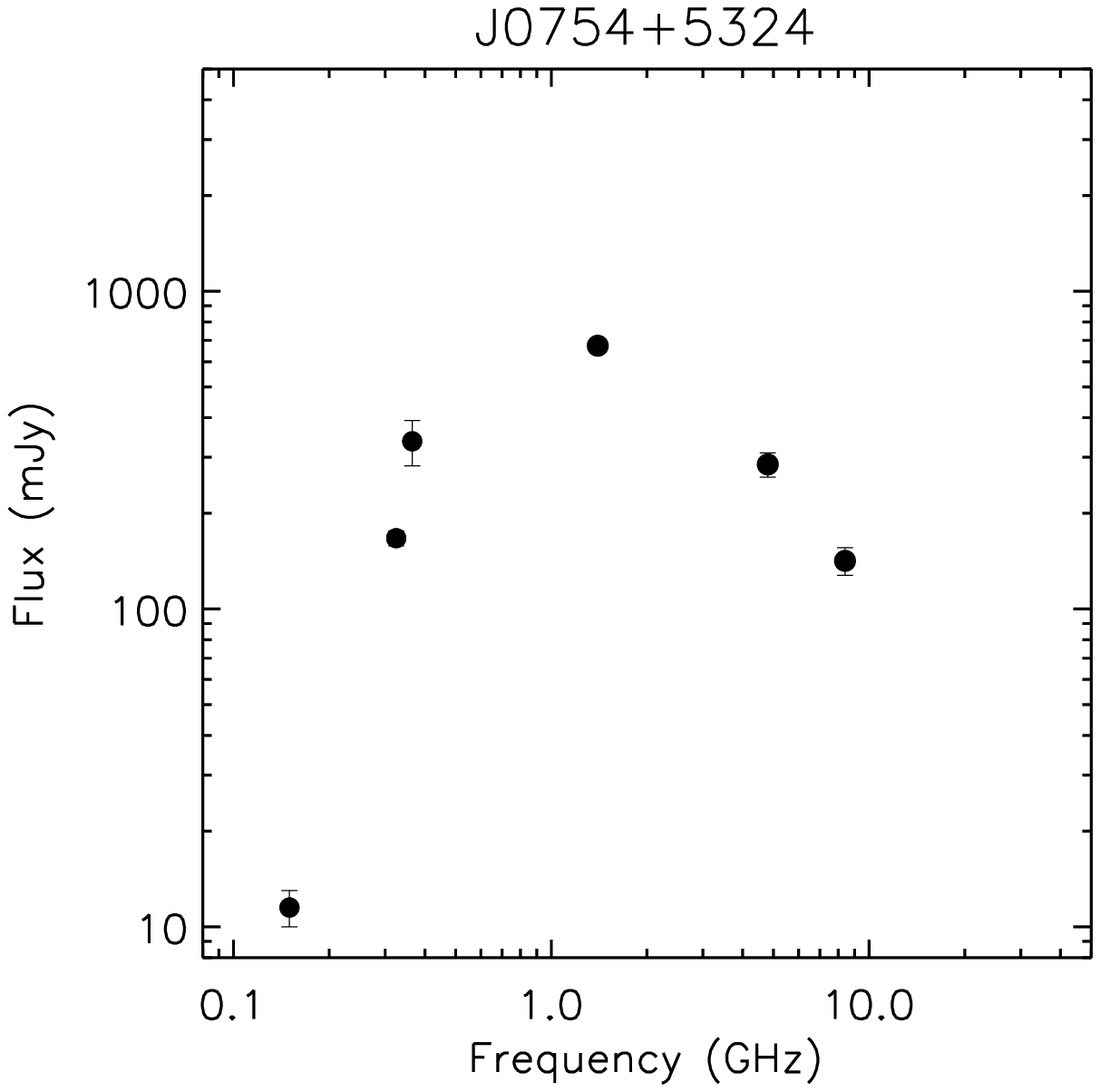} 
\includegraphics[scale=.45]{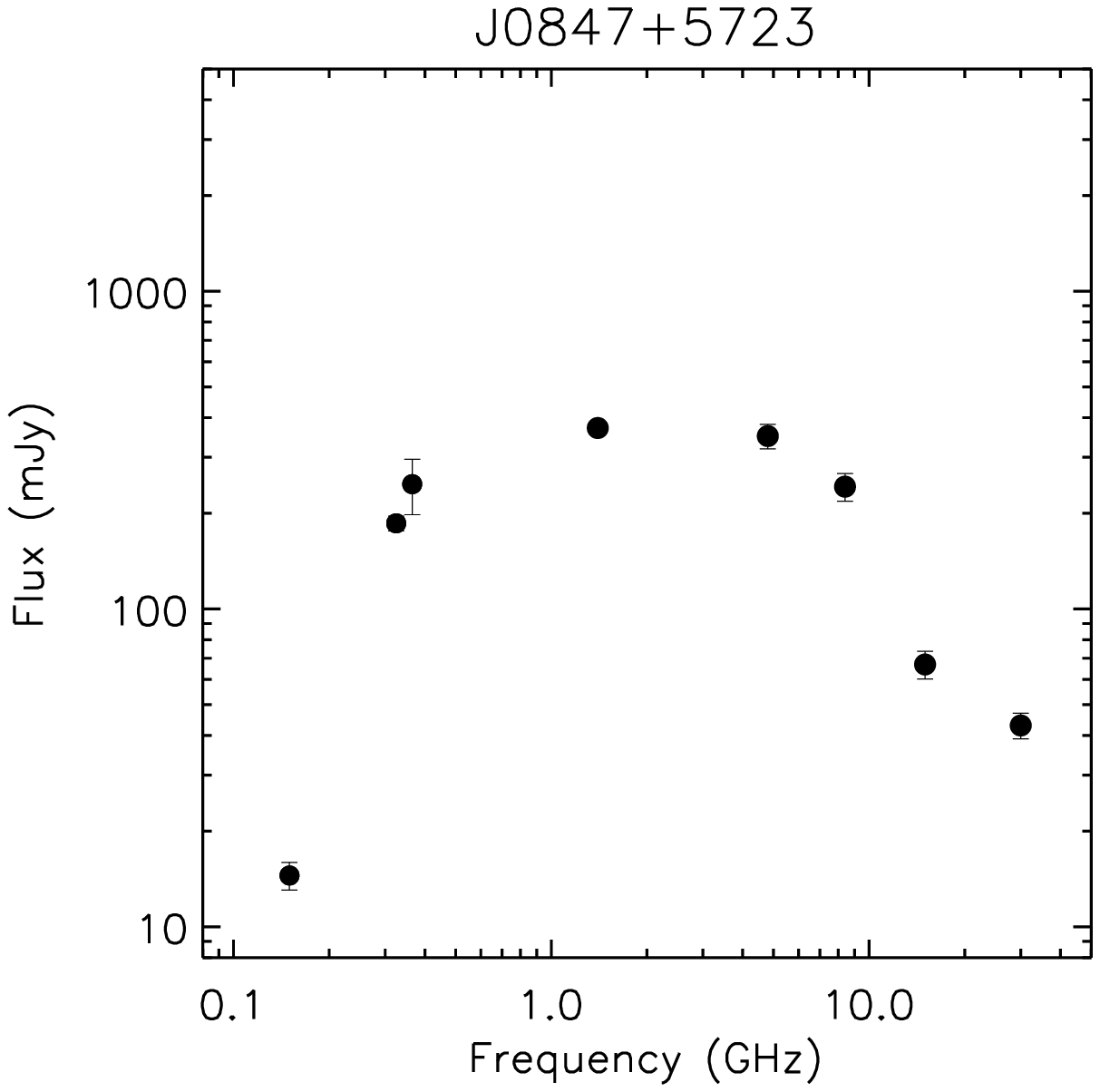} 
\includegraphics[scale=.45]{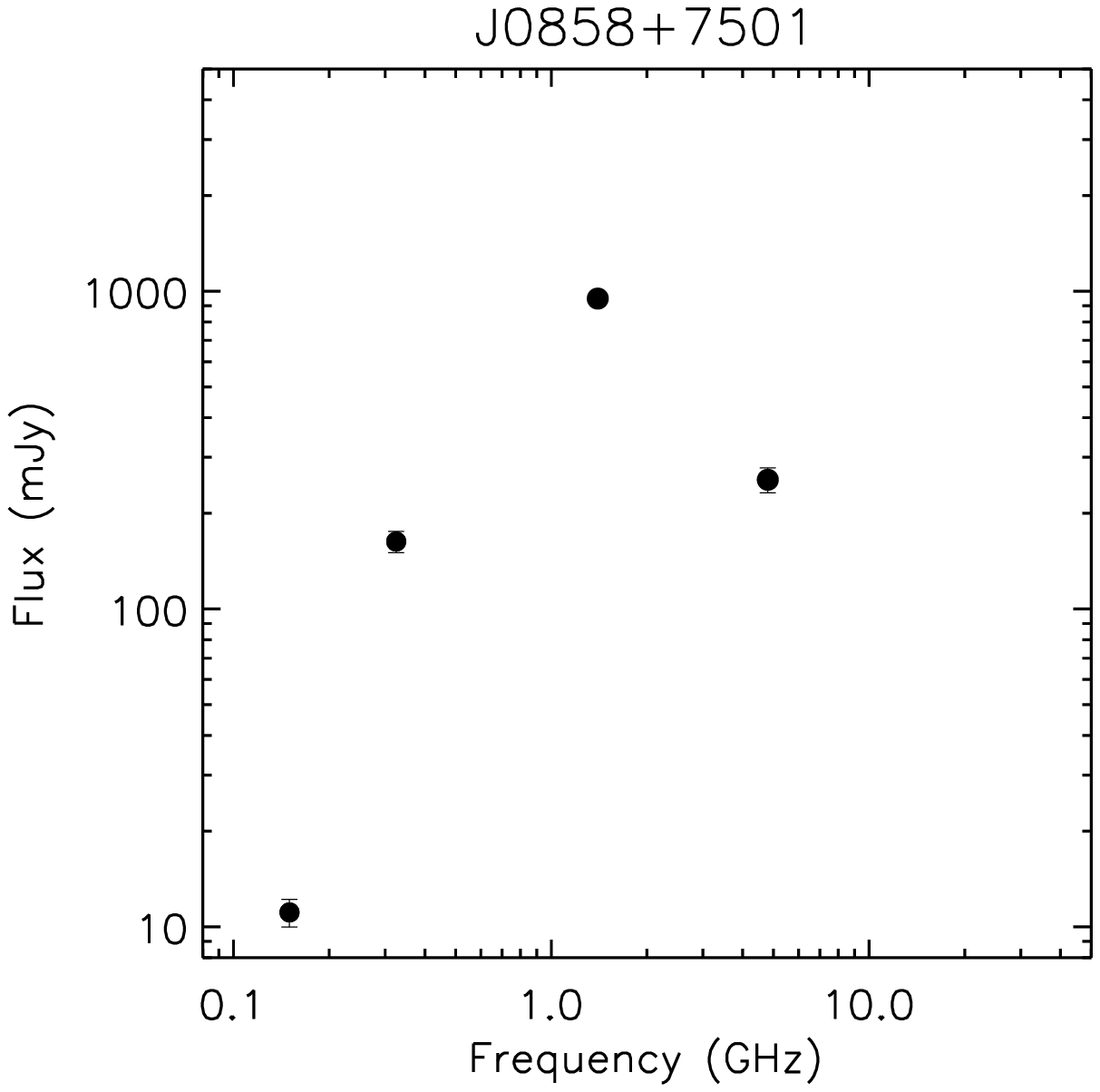} 
\includegraphics[scale=.45]{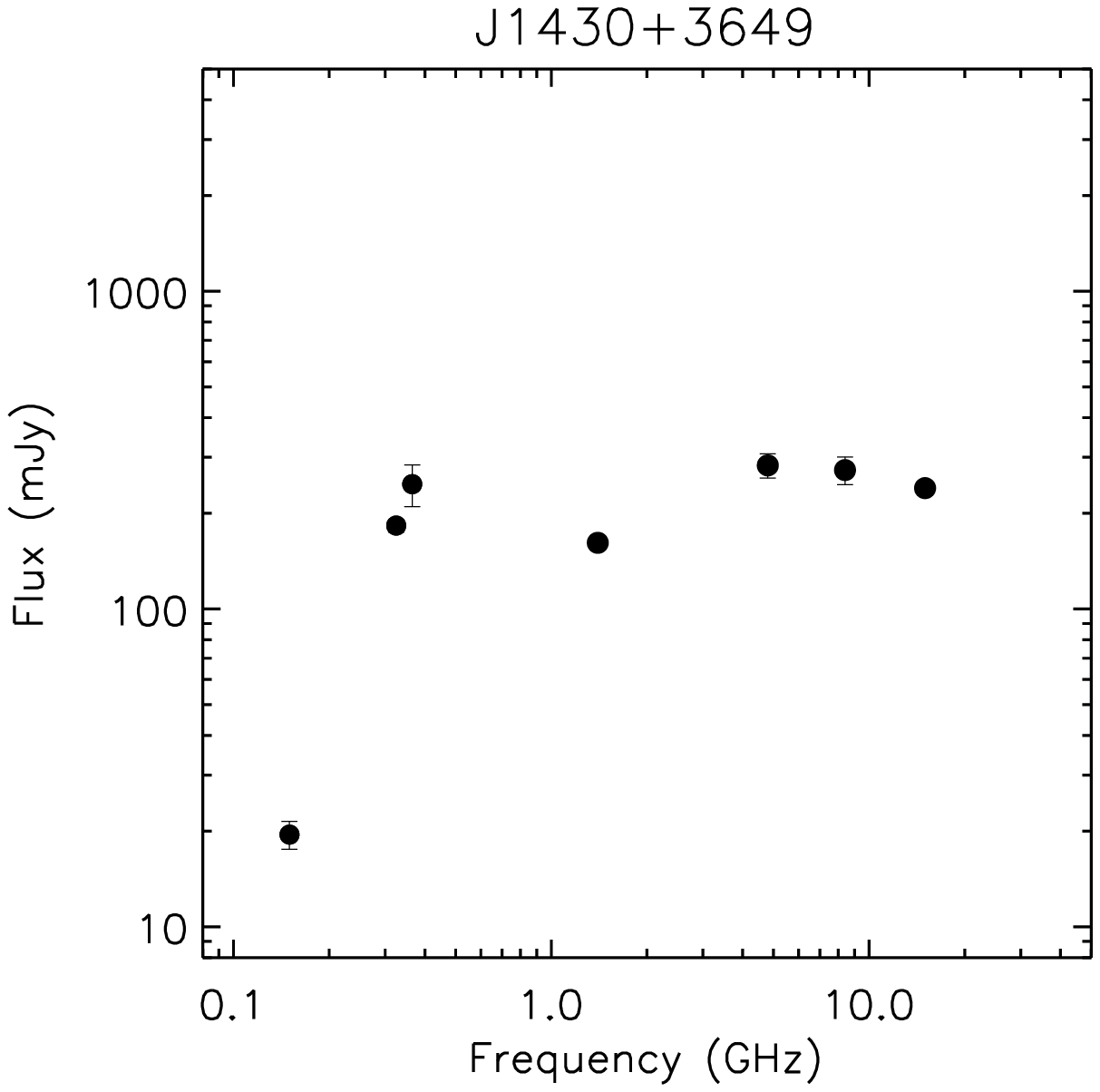} 
\includegraphics[scale=.45]{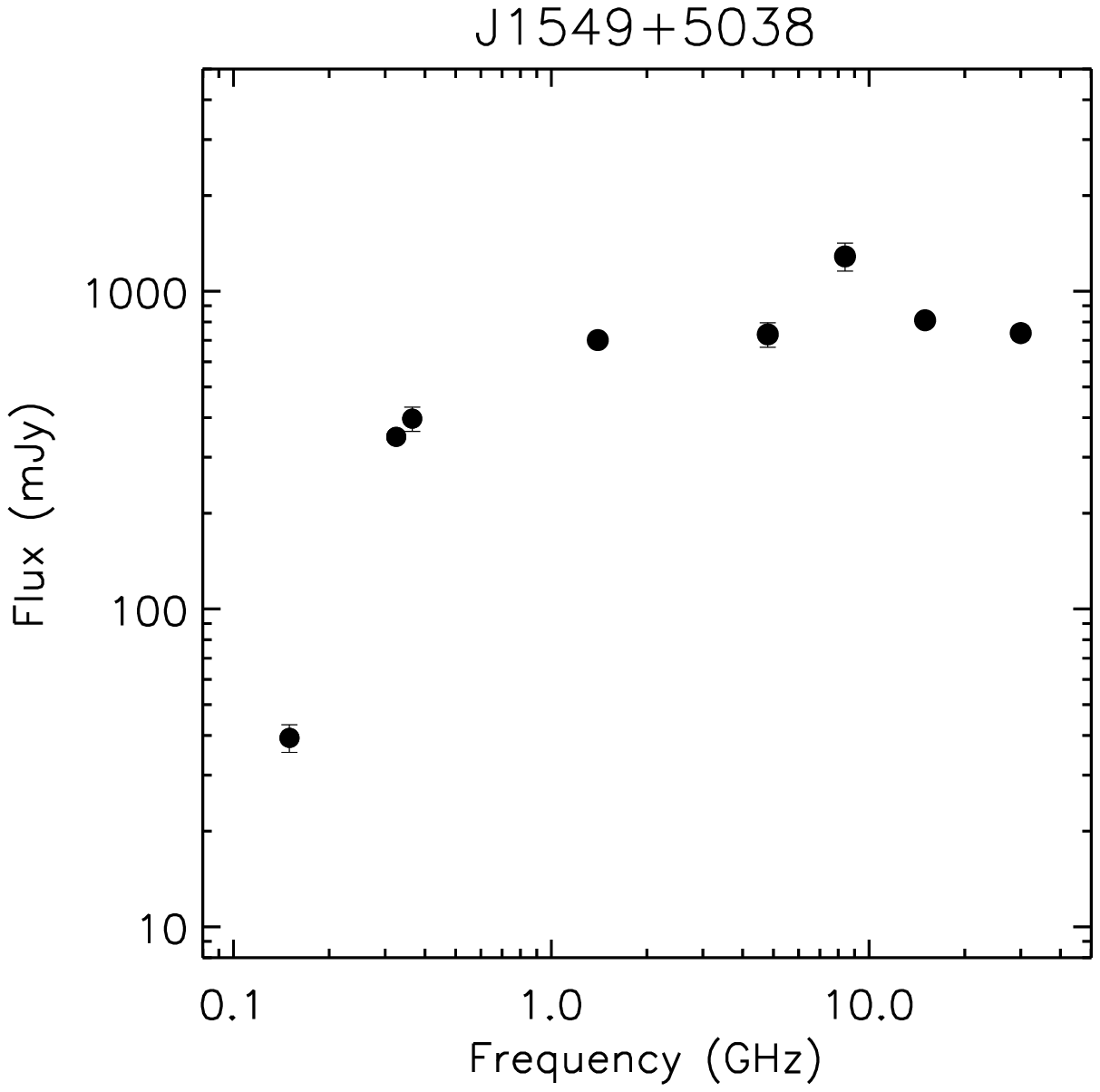} 
\includegraphics[scale=.45]{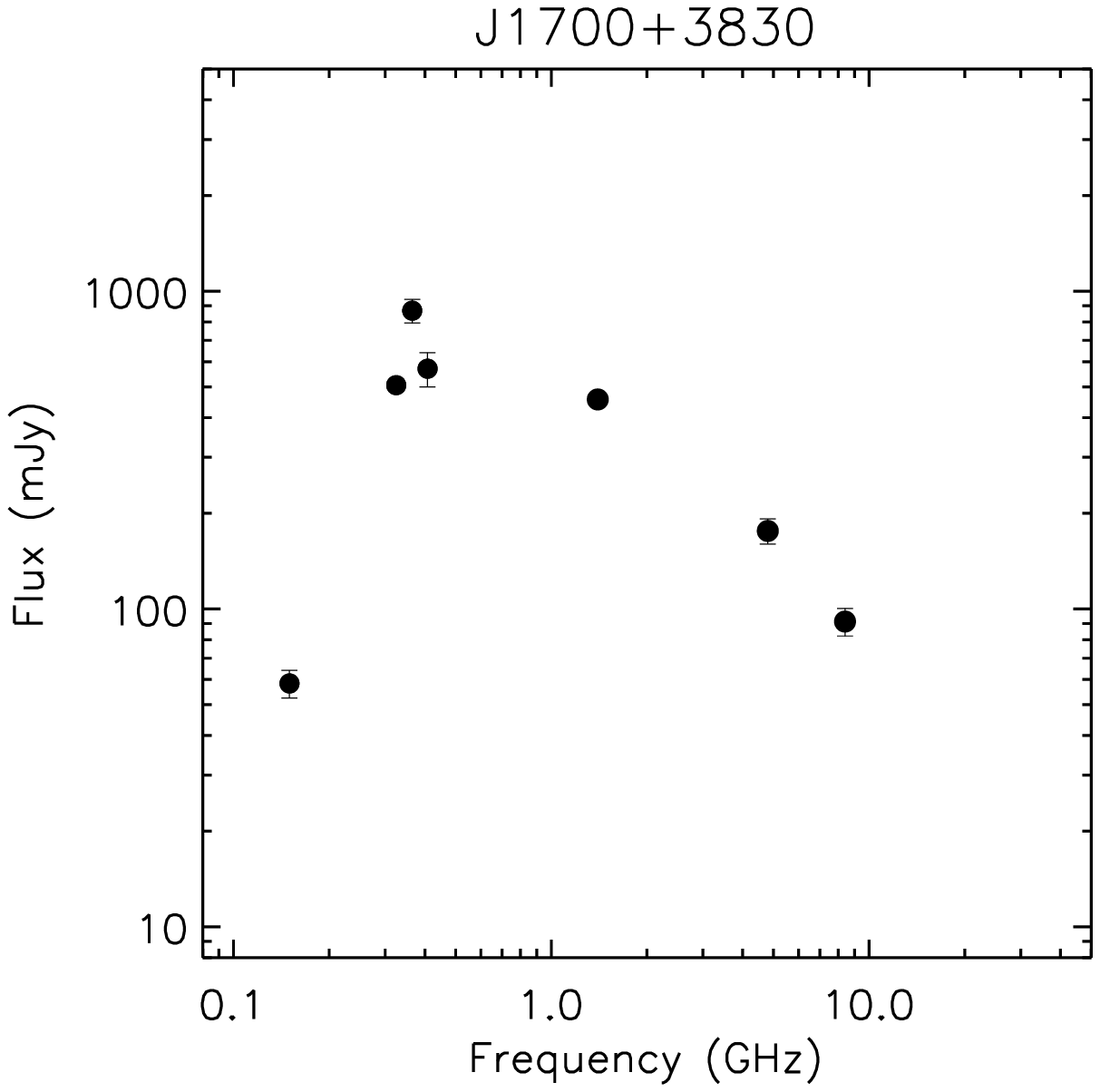} 
\includegraphics[scale=.45]{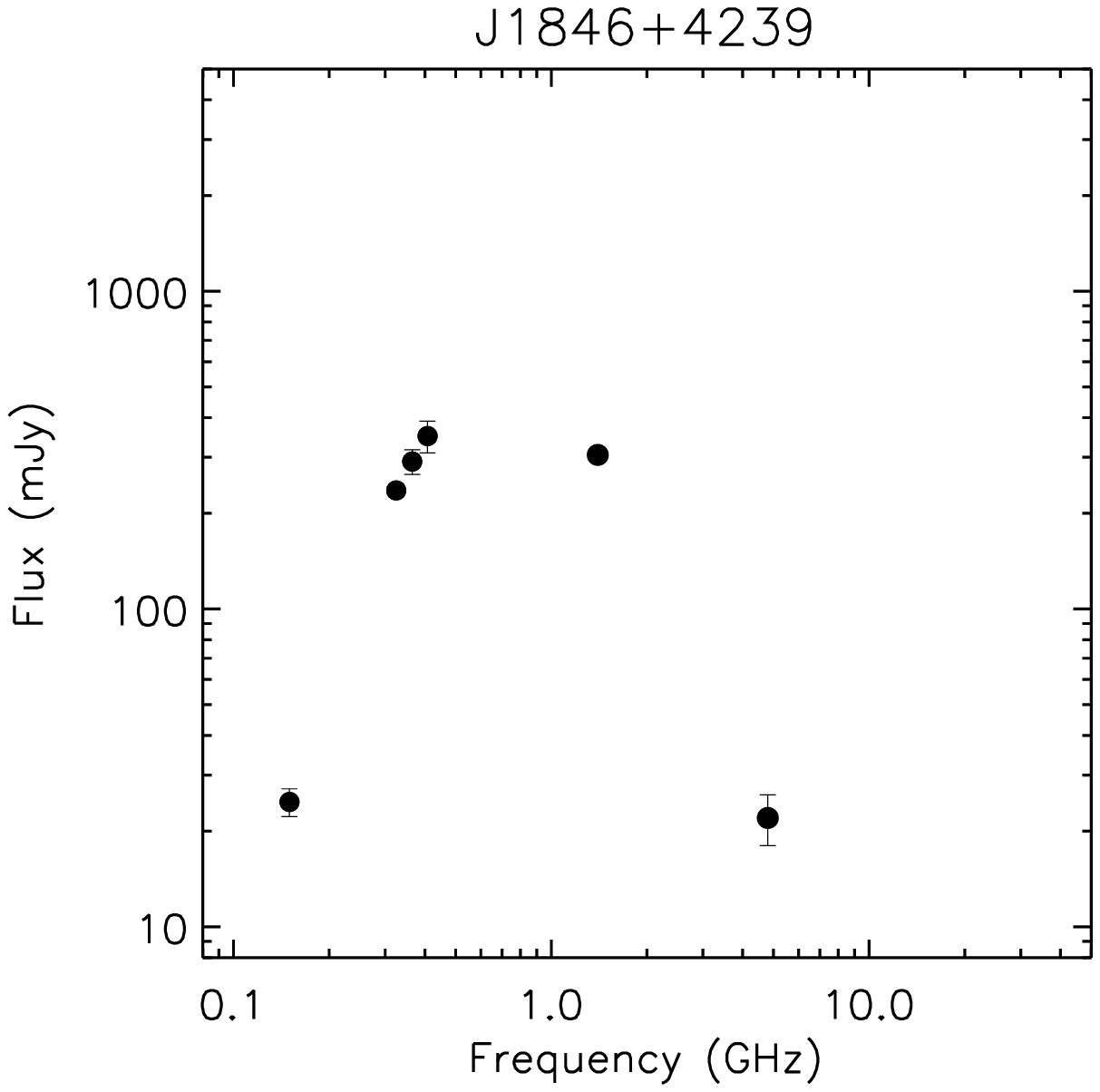}
\caption{\scriptsize{Radio spectra of the 7 EISERS candidates detected in the 150 MHz ADR-TGSS, based on non-contemporaneous flux measurements (see Table ~\ref{table:spec-prop}).}}
\label{fig:spec_all_detected}
\end{figure}

\begin{figure}[!htbp]
\centering
\includegraphics[scale=.45]{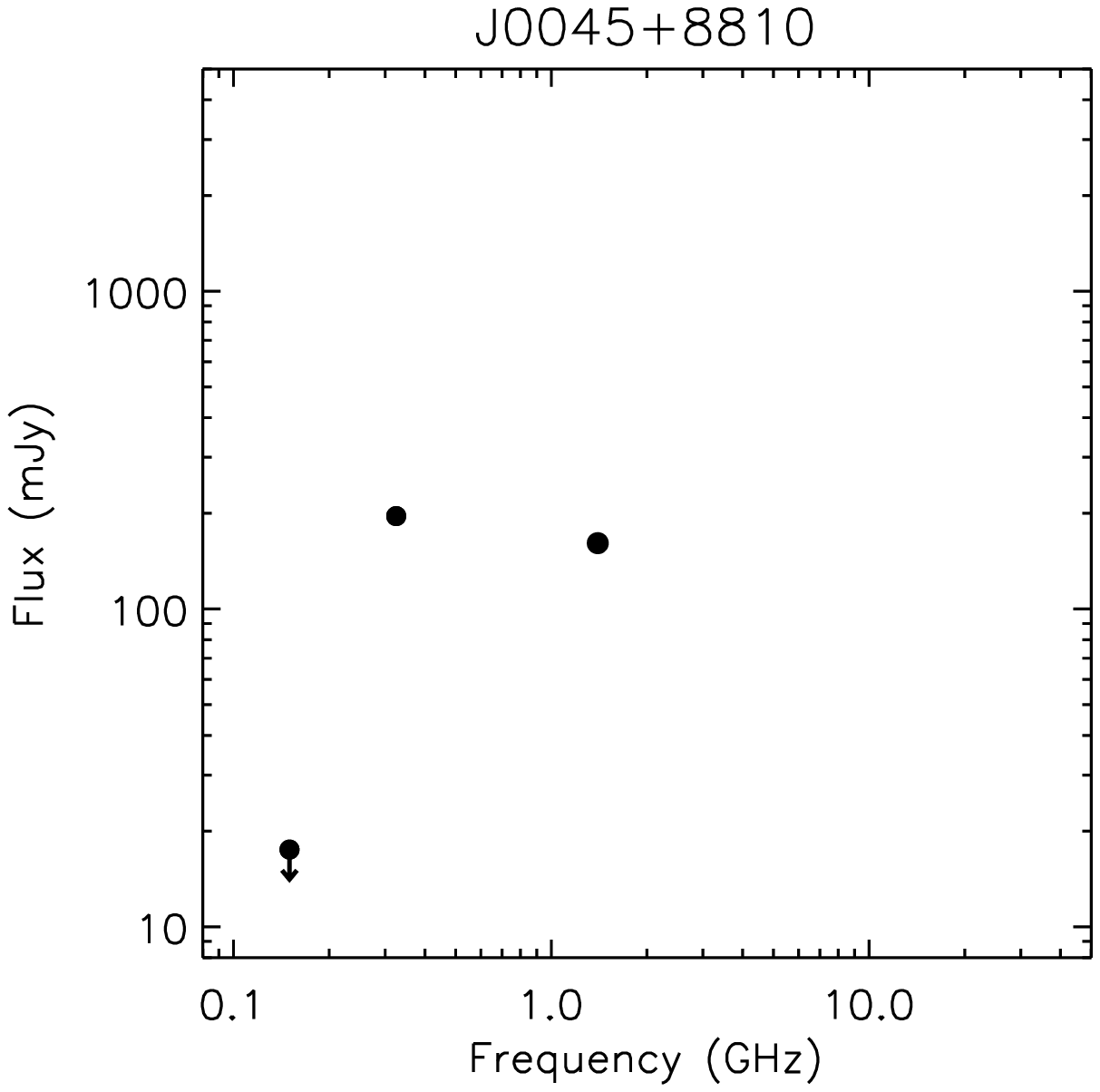} 
\includegraphics[scale=.45]{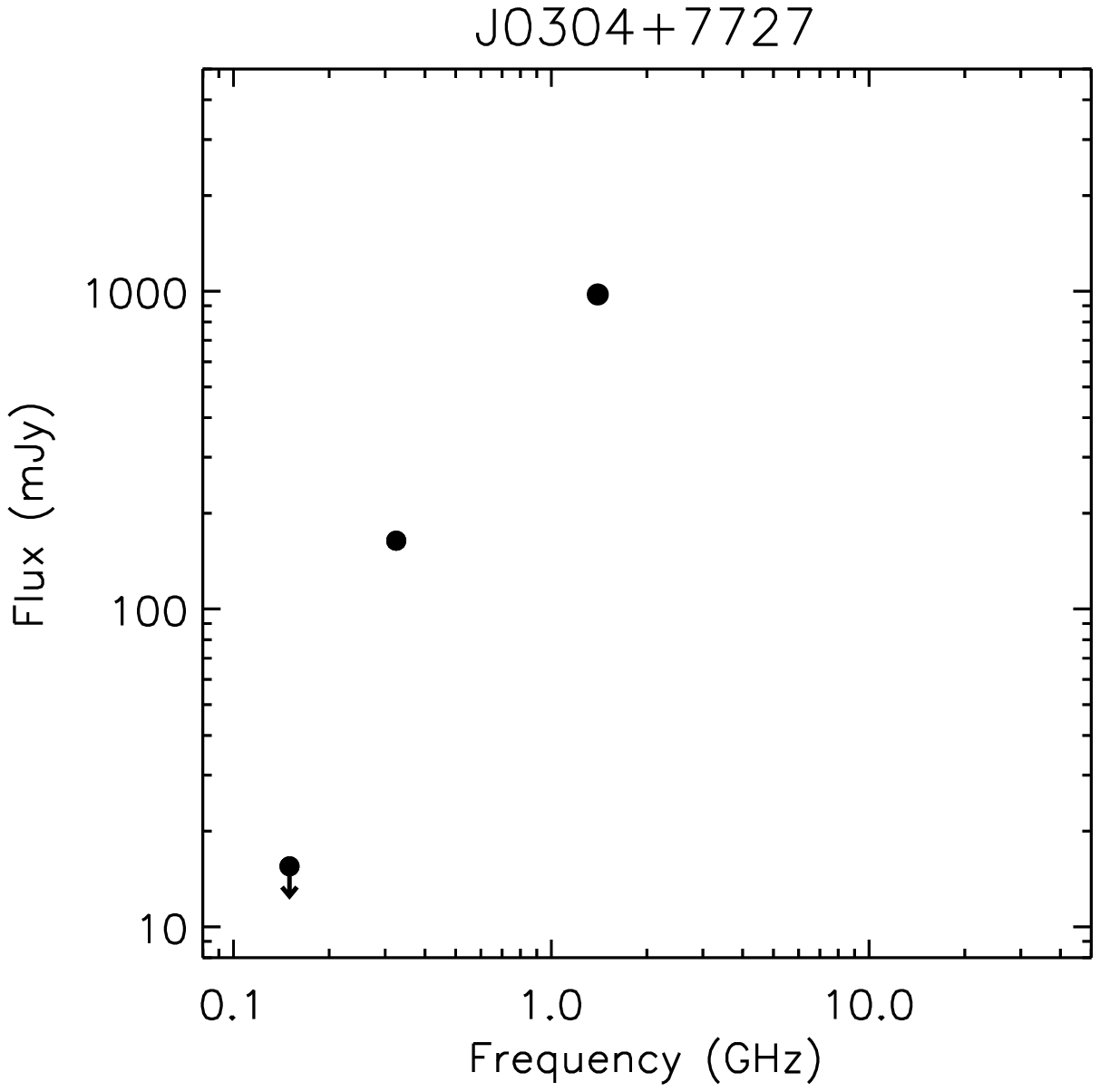}  
\includegraphics[scale=.45]{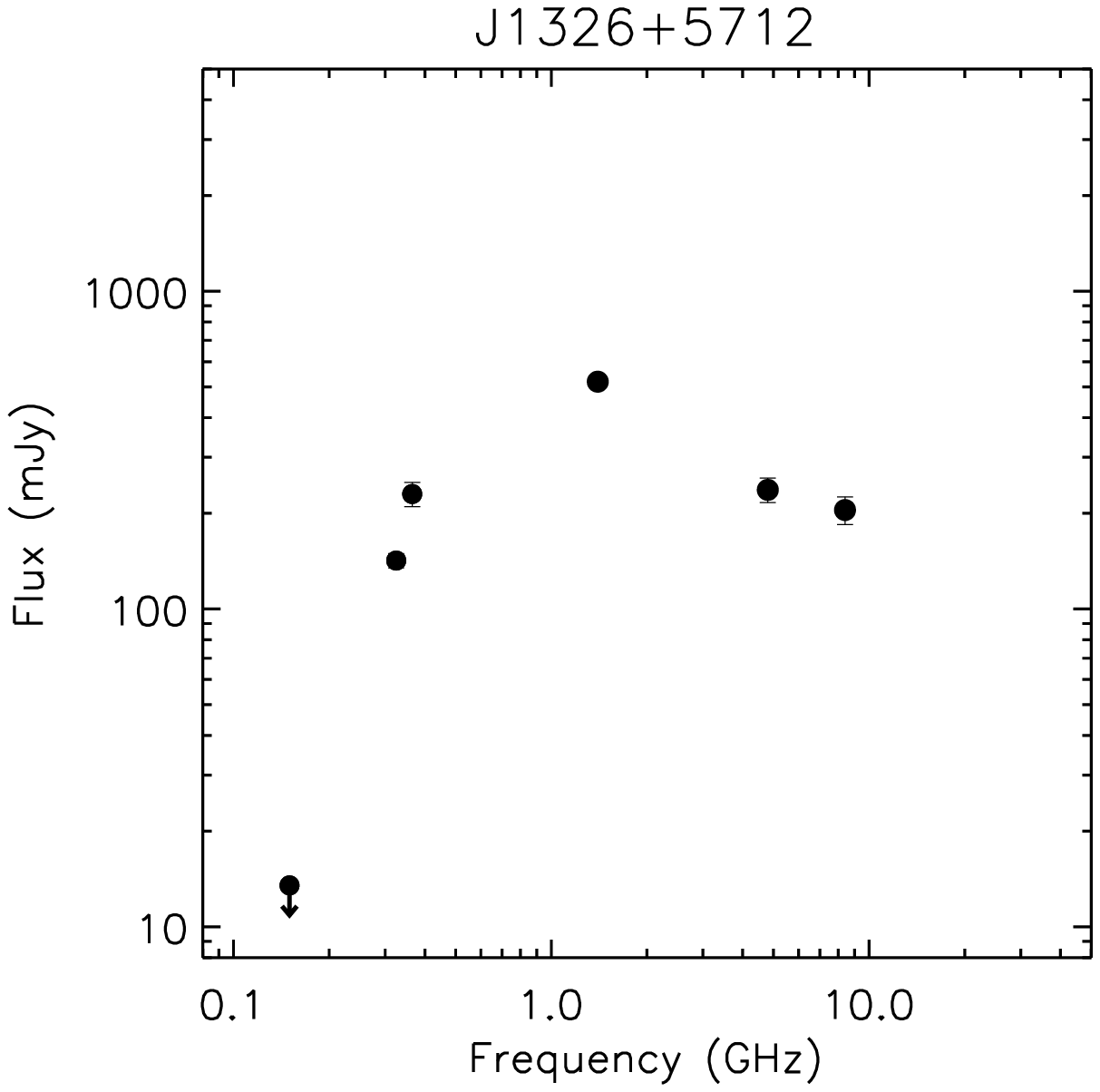} 
\includegraphics[scale=.45]{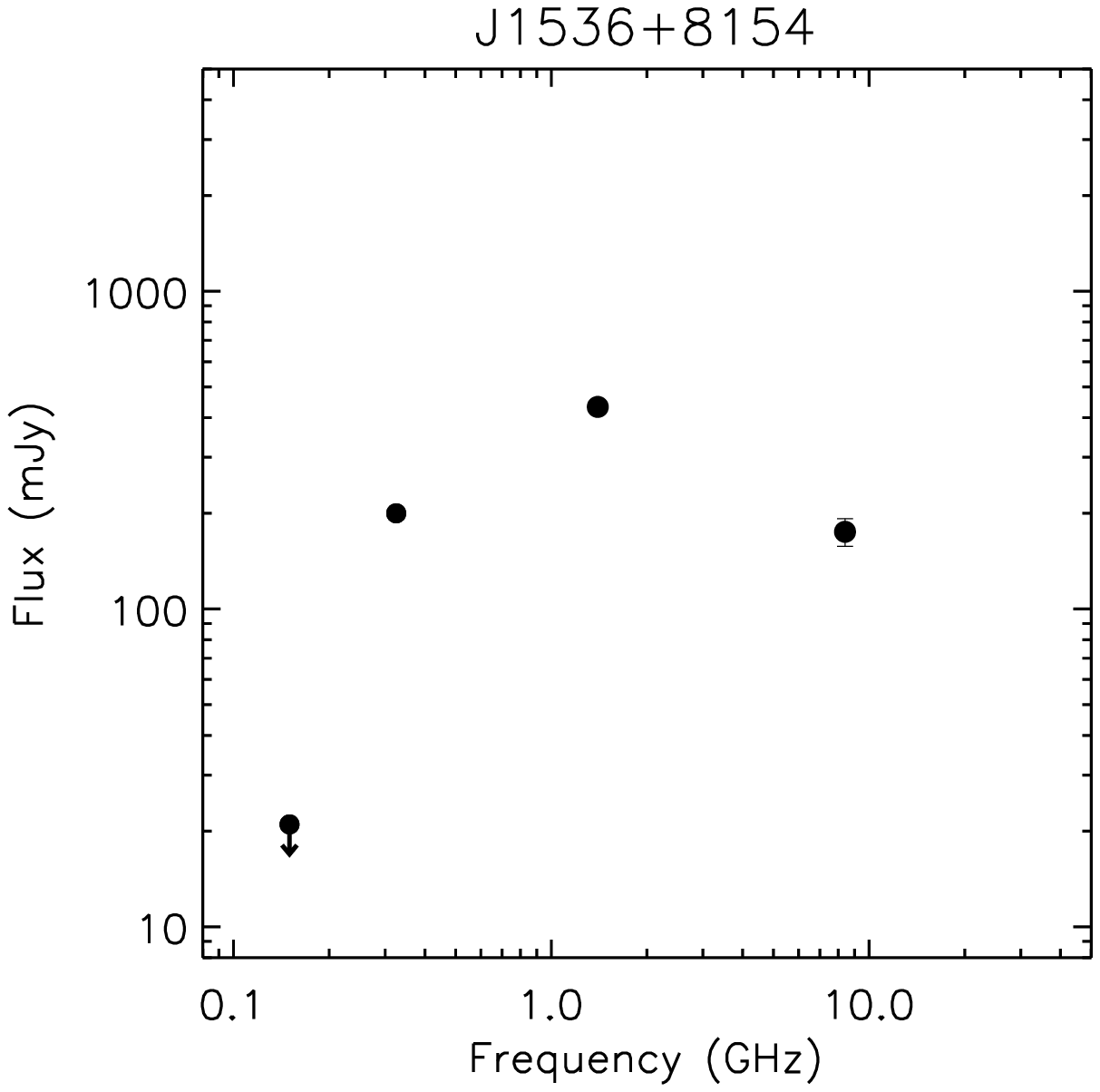} 
\includegraphics[scale=.45]{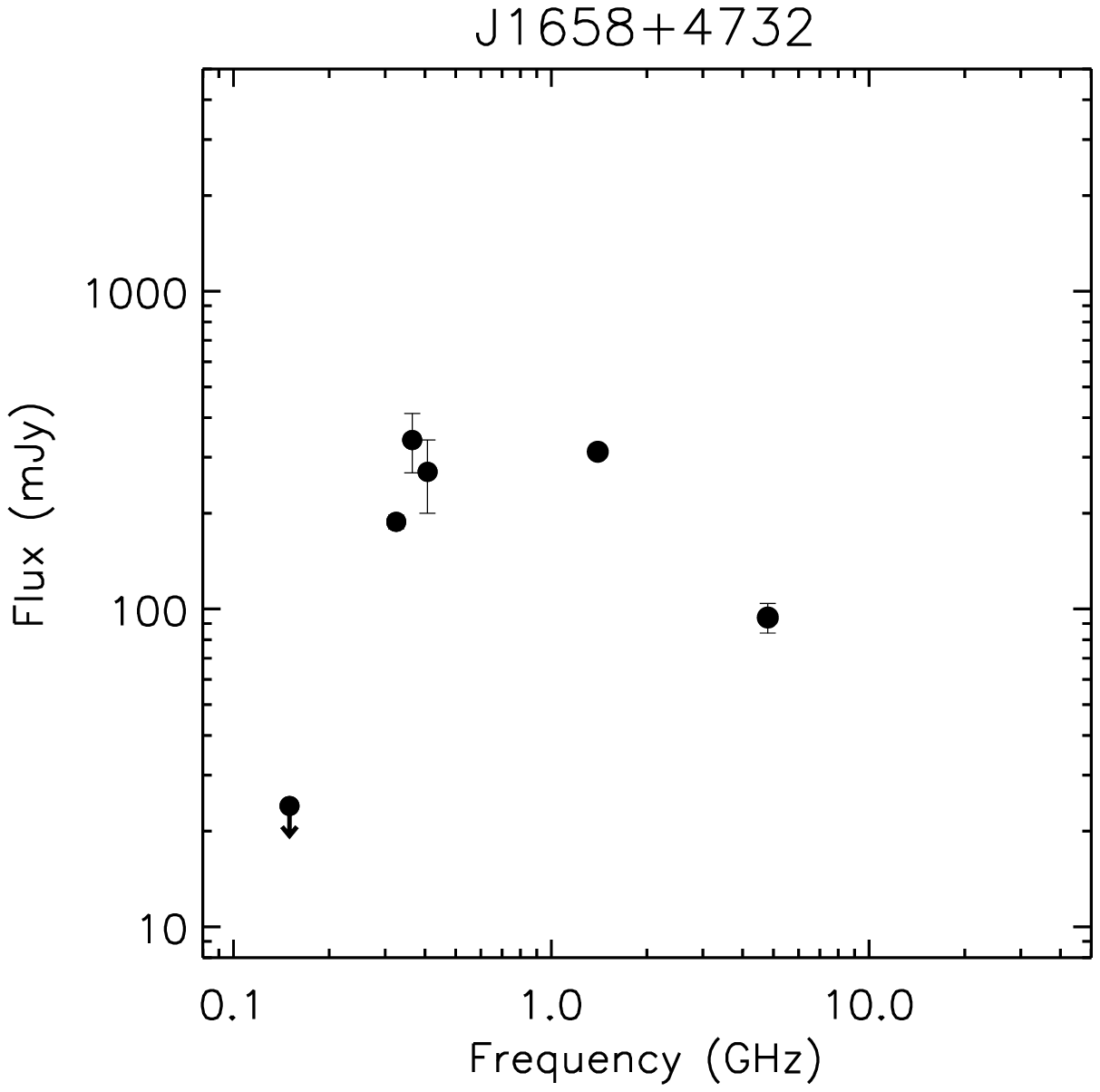} 
\includegraphics[scale=.45]{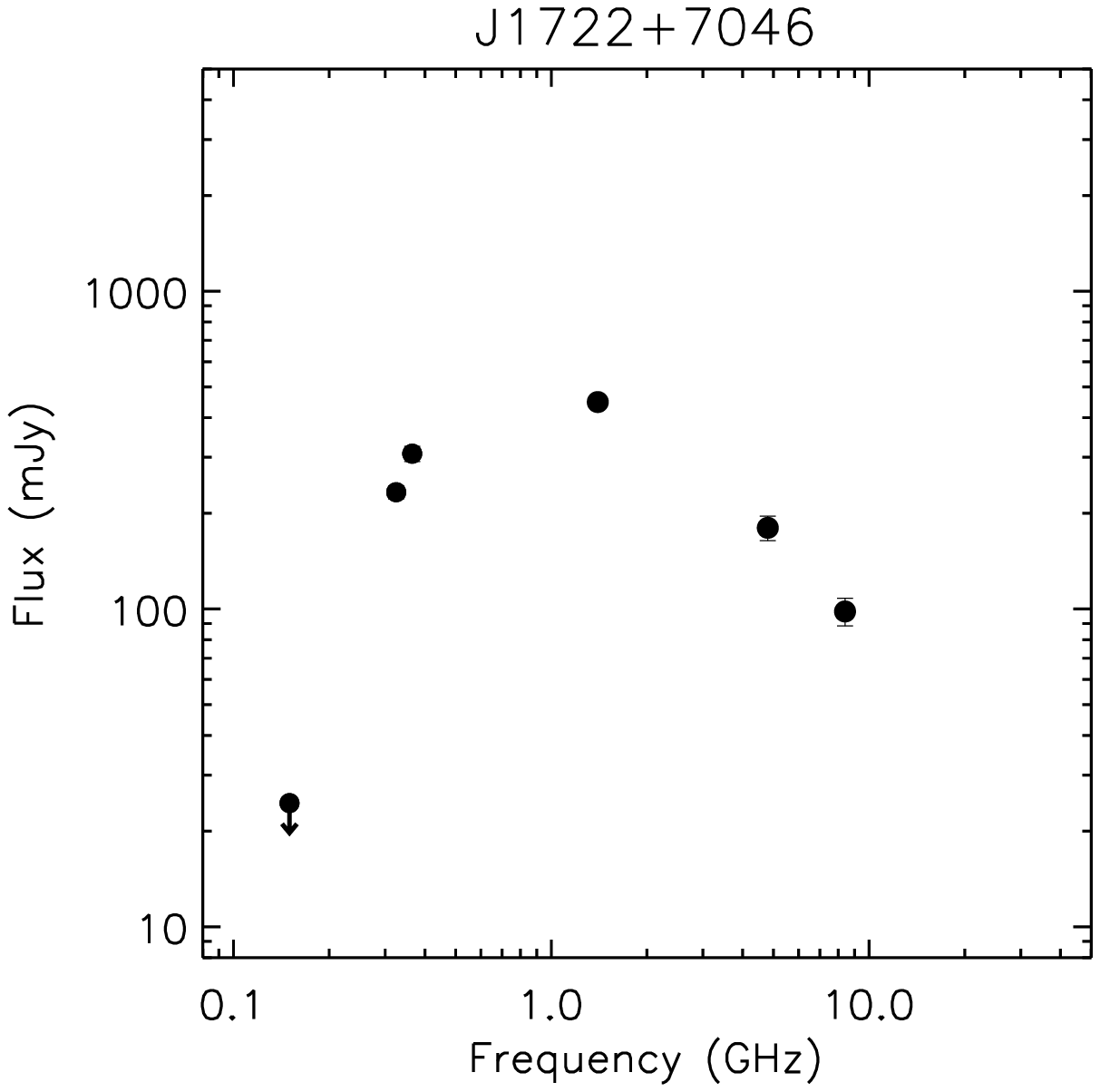} 
\includegraphics[scale=.45]{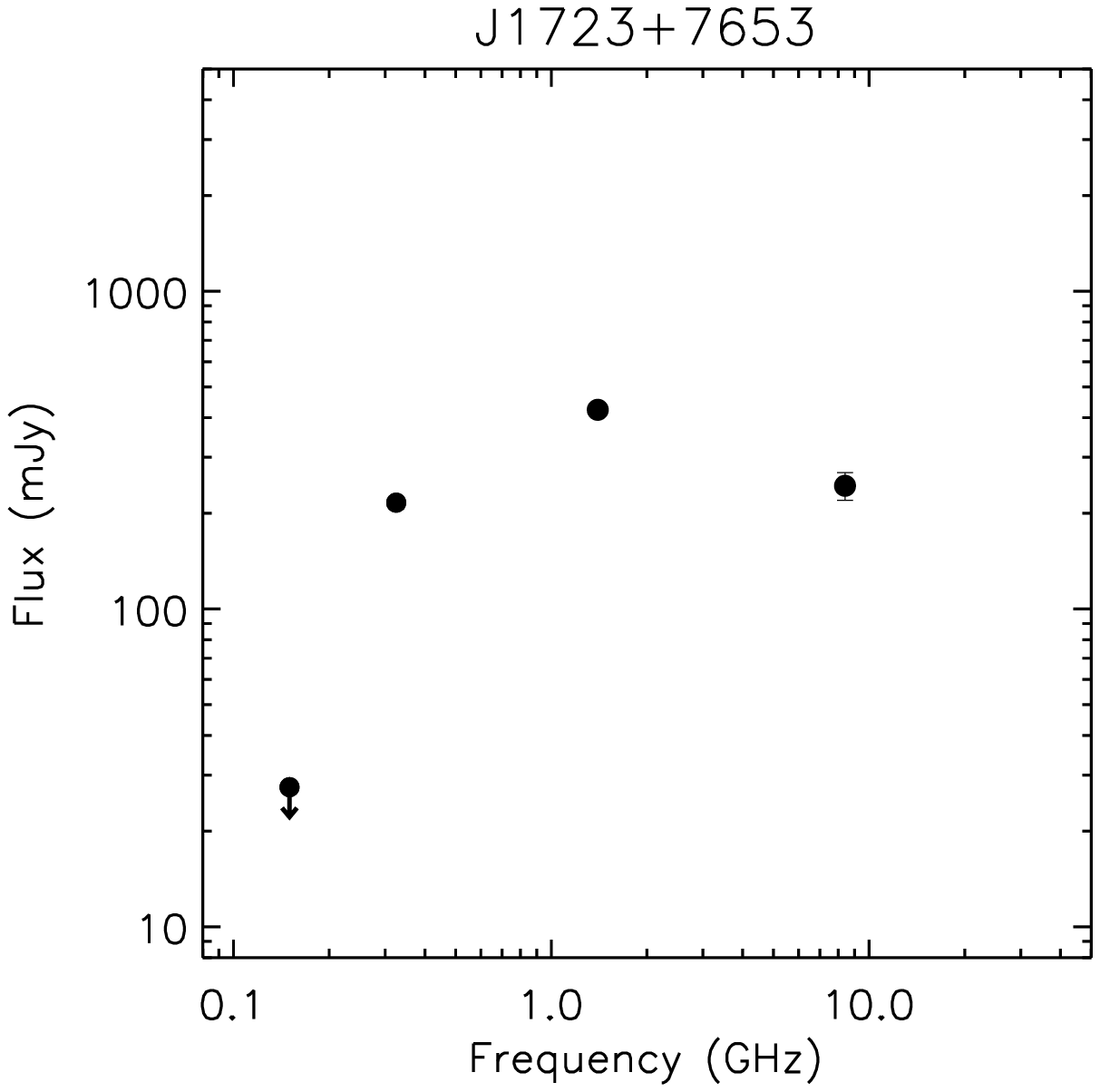} 
\includegraphics[scale=.45]{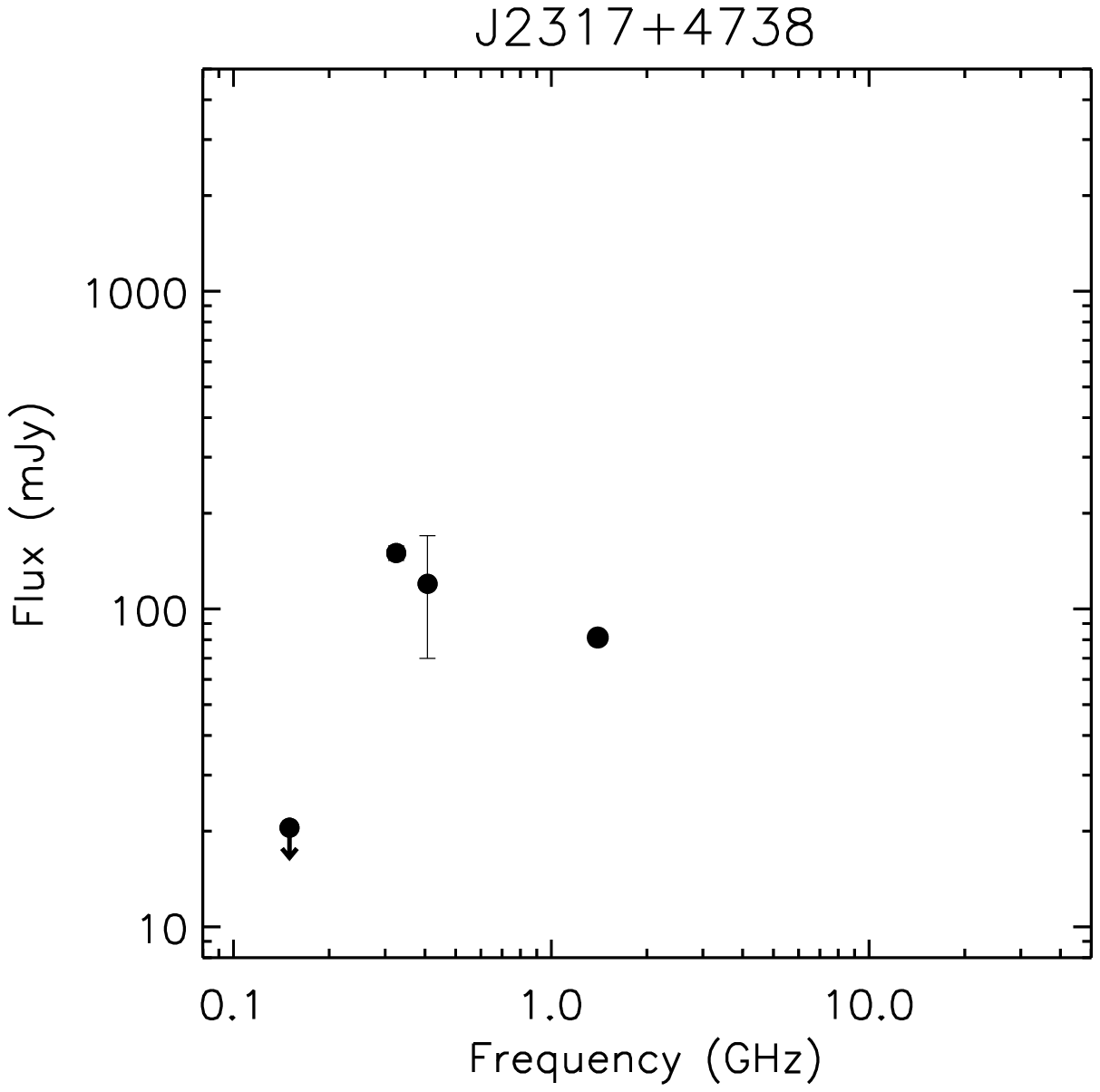}
\caption{\scriptsize{Radio spectra of the 8 EISERS candidates un-detected in the 150 MHz ADR-TGSS, based on non-contemporaneous flux measurements (see Table ~\ref{table:spec-prop}).}}
\label{fig:spec_all_undetected}
\end{figure}

The above sequence of steps led to a list of 72 sources (including 36 for which only a lower limit to $\alpha$ (150-325 MHz) could be placed) satisfying the following criteria: (i) Structural type `S' and flux density $>$ 150 mJy at 325 MHz, as per the WENSS catalogue (ii) $\alpha$ (150-325 MHz) $>$ +2.0 (iii) rejection of sources that are termed extended in either of the two basic survey catalogues, or lie in the galactic plane, or are associated with known H-II regions and (iv) a clean detection in at least the WENSS (325 MHz) and NVSS (1.4 GHz) (based on visual inspection of the respective radio images). Finally, we extracted out of these 72 sources, our final list by demanding that the estimated $\alpha$ (150-325 MHz) of the source exceeds +2.75, or its lower limit exceeds +2.50. The final list of 15 EISERS candidates is presented in Table~\ref{table:spec-prop}, after summarizing the sequence of the selection steps in Table~\ref{table:numbers}. It needs to be reiterated that these estimates of $\alpha$ (150-325 MHz) are based on non-contemporaneous flux density measurements and should, strictly speaking, be treated as tentative. Table~\ref{table:opt-prop} lists some salient observational parameters of the 15 EISERS candidates. Their radio spectra, based on the data provided in Table~\ref{table:spec-prop} and comments in Sect. 3, are displayed in Figs~\ref{fig:spec_all_detected} \& ~\ref{fig:spec_all_undetected}.

\begin{table*}
\small
\centering
\caption{Number of sources filtered at the successive stages of the selection process.}
\label{table:numbers}
\begin{tabular}{llc}\\
\hline
\multicolumn{1}{l}{Occurrence in the catalogues} & \multicolumn{1}{l}{Imposed criteria} & \multicolumn{1}{l}{Number of sources}\\
\multicolumn{1}{l}{} & \multicolumn{1}{l}{} & \multicolumn{1}{l}{satisfying the criteria}\\
\hline
WENSS only &S-type, $>$150 mJy & 35064\\ 
WENSS \& ADR-TGSS  & S-type, $>$150 mJy & 33707\\
WENSS but NOT in ADR-TGSS  & S-type, $>$150 mJy & 1357\\
WENSS \& ADR-TGSS  & S-type, $>$150 mJy, $\alpha$ $>$ +2.00 & 36\\
WENSS but NOT in ADR-TGSS  & S-type, $>$150 mJy, $\alpha$ $>$ +2.00 & 36\\
WENSS \& ADR-TGSS  & S-type, $>$150 mJy, $\alpha$ $=$ +2.75 or more & 07\\
WENSS but NOT in ADR-TGSS  & S-type, $>$150 mJy, $\alpha$ (lower limit)$=$ +2.50 & 08\\
\hline
\end{tabular}
\end{table*}

\begin{table}
\small
\addtolength{\tabcolsep}{-3pt}
\caption{Positions, flux densities and spectral indices (150-325 MHz) of the 15 EISERS candidates.}
\label{table:spec-prop}
\begin{tabular}{cccllllllc}\\
\hline

\multicolumn{1}{c}{Source Name} & \multicolumn{1}{c}{RA} & \multicolumn{1}{c}{DEC}& \multicolumn{1}{l}{150 MHz}& \multicolumn{1}{l}{325 MHz}& \multicolumn{1}{l}{1400 MHz}& \multicolumn{1}{l}{4.85 GHz} & \multicolumn{1}{l}{8.4 GHz} &\multicolumn{1}{l}{Spectral} & \multicolumn{1}{c}{Notes}\\
  &(J2000)  &(J2000)& ADR-TGSS$^{\dagger}$ & WENSS$^{\dagger}$ & FIRST$^{\dagger}$ &PMN$^{\dagger}$ & CLASS$^{\dagger}$ & Index &  \\
  &    & & (mJy)                &(mJy)              &(mJy)             &(mJy)           &(mJy)             &$\alpha$       &  \\

\hline
\\
J0754+5324& 07 54 15.36&+53 24 56.52&11.5${\pm}$1.2& 167.0${\pm}$09.0& 676.3${\pm}$0.23& 285.0${\pm}$25.0&  141.7${\pm}$14.1&3.46${\pm}$0.15&$^{\star\star}$\\
J0847+5723& 08 47 27.86&+57 23 35.52&14.5${\pm}$1.5& 186.0${\pm}$10.0& 363.2${\pm}$0.14& 350.0${\pm}$31.0&  242.6${\pm}$24.2&3.30${\pm}$0.15&$^{\star\star}$\\
J0858+7501& 08 58 33.43&+75 01 18.47&11.1${\pm}$1.1& 163.0${\pm}$12.6& 947.4${\pm}$28.4$^{\ast}$& 255.0${\pm}$23.0&		   &3.47${\pm}$0.16&\\
J1430+3649& 14 30 40.63&+36 49 07.32&19.5${\pm}$2.0& 183.0${\pm}$08.6& 162.0${\pm}$0.15& 283.0${\pm}$25.0&  273.5${\pm}$27.3&2.90${\pm}$0.14&$^{\star\star}$\\
J1549+5038& 15 49 17.46&+50 38 05.32&39.3${\pm}$3.9& 348.0${\pm}$14.6& 701.4${\pm}$0.14& 731.0${\pm}$65.0& 1286.1${\pm}$128.6&2.82${\pm}$0.14&$^{\star\star}$\\
J1700+3830& 17 00 19.96&+38 30 33.81&58.3${\pm}$5.8& 506.0${\pm}$20.9& 430.2${\pm}$0.14& 176.0${\pm}$16.0&   091.3${\pm}$9.1&2.79${\pm}$0.14&$^{\star\star}$\\
J1846+4239& 18 46 42.63&+42 39 45.68&24.7${\pm}$2.5& 236.0${\pm}$10.4& 305.3${\pm}$09.2$^{\ast}$&  022.0${\pm}$04.0&		   &2.92${\pm}$0.14&$^{\star\star}$\\
\\
\multicolumn{9}{c}{The 8 EISERS candidates with only a lower limit to $\alpha$ (150-325 MHz)}\\
\\
J0045+8810& 00 45 05.38&+88 10 19.19&$<$17.5	    & 196.0${\pm}$08.3& 161.1${\pm}$04.8$^{\ast}$&		  &		   &$>$3.12&$^{\star\star}$\\
J0304+7727& 03 04 55.06&+77 27 31.68&$<$15.5       & 164.0${\pm}$07.2& 976.3${\pm}$29.3$^{\ast}$&		  &		   &$>$3.05&\\
J1326+5712& 13 26 50.47&+57 12 06.84&$<$13.5       & 142.0${\pm}$07.4& 528.5${\pm}$0.16& 237.0${\pm}$21.0&  204.8${\pm}$20.4&$>$3.04&$^{\star\star}$\\
J1536+8154& 15 36 59.83&+81 54 31.32&$<$21.0       & 200.0${\pm}$08.5& 432.3${\pm}$13.0$^{\ast}$&		  &  174.9${\pm}$17.4&$>$2.91&$^{\star\star}$\\
J1658+4732& 16 58 26.54&+47 32 14.99&$<$24.0       & 188.0${\pm}$09.3& 304.6${\pm}$0.18&  094.0${\pm}$10.0&		   &$>$2.66&$^{\star\star}$\\
J1722+7046& 17 22 07.22&+70 46 28.19&$<$24.5       & 233.0${\pm}$11.2& 447.8${\pm}$13.4$^{\ast}$& 180.0${\pm}$16.0&   098.2${\pm}$9.8&$>$2.91&$^{\star\star}$\\
J1723+7653& 17 23 59.88&+76 53 11.75&$<$27.5       & 216.0${\pm}$09.1& 423.3${\pm}$12.7$^{\ast}$&		  &  244.1${\pm}$24.4&$>$2.67&$^{\star\star}$\\
J2317+4738& 23 17 10.83&+47 38 21.48&$<$20.5       & 150.0${\pm}$07.9& 081.3${\pm}$02.5$^{\ast}$&		  &		   &$>$2.57&\\

\hline
\end{tabular}

{$^{\dagger}$ ADR-TGSS--\citet{Intema2017}; WENSS -- \citet{deBruyn2000}; FIRST -- \citet{Helfand2015}; \\ 
PMN -- \citet{Griffith1994}; CLASS -- \citet{Myers2003}; NED -- NASA Extragalactic Database.}\\
{$^{\ast}$ Outside the FIRST Survey range, hence flux density taken from the NVSS \citep{Condon1998}.}\\
{$^{\star\star}$Possible counterpart to a $\gamma$-ray source \citep{Massaro2014}}
\end{table}

\begin{table}
\addtolength{\tabcolsep}{-3pt}
\caption{Optical and Infrared properties of the 15 EISERS candidates.}
\label{table:opt-prop}
\begin{tabular}{cccccccccccccc}\\
\hline
  
\multicolumn{1}{c}{Source} & \multicolumn{1}{c}{Opt ID$^{\dagger}$} & \multicolumn{1}{c}{z$^{\dagger}$} & \multicolumn{1}{c}{u} & \multicolumn{1}{c}{g} & \multicolumn{1}{c}{r}& \multicolumn{1}{c}{i}& \multicolumn{1}{c}{z}& \multicolumn{1}{c}{3.4 $\mu$m} & \multicolumn{1}{c}{4.6$\mu$m} &\multicolumn{1}{c}{12$\mu$m} & \multicolumn{1}{c}{22 $\mu$m} & \multicolumn{1}{c}{u-g} & \multicolumn{1}{c}{u-r}\\
  & & &mag  &mag &mag &mag &mag &mag &mag &mag &mag &mag &mag  \\
  &   & &SDSS$^{\ddagger}$ &SDSS$^{\ddagger}$ &SDSS$^{\ddagger}$ &SDSS$^{\ddagger}$ &SDSS$^{\ddagger}$&WISE$^{\ddagger}$  &WISE$^{\ddagger}$    &WISE$^{\ddagger}$    &WISE$^{\ddagger}$    &    & \\

\hline
\\
J0754+5324	&	QSO	&		&	18.22	&	16.65	&	16.07	&	15.87	&	15.79	&	16.802	&	15.24	&	11.641	&	$>$8.782	&	1.57	&	2.15	\\
J0847+5723	&	QSO	&		&	24.31	&	23.67	&	22.12	&	22.28	&	21.95	&	15.71	&	14.897	&	10.665	&	8.523	&	0.64	&	2.19	\\
J0858+7501	&	QSO	&		&		&		&		&		&		&	16.715	&	15.823	&	11.757	&	8.265	&		&		\\
J1430+3649	&	QSO	&	2.608	&	20.22	&	19.61	&	19.17	&	18.74	&	18.4	&	14.614	&	13.737	&	10.973	&	8.75	&	0.61	&	1.05	\\
J1549+5038	&	QSO	&	2.170	&	18.82	&	18.48	&	18.52	&	18.36	&	18.16	&	14.951	&	13.93	&	10.694	&	8.161	&	0.34	&	0.3	\\
J1700+3830	&	QSO	&		&	24.42	&	22.57	&	21.56	&	20.57	&	19.93	&		&		&		&		&	1.85	&	2.86	\\
J1846+4239	&		&		&		&		&		&		&		&	15.506	&	15.406	&	$>$13.243	&	$>$9.093	&		&		\\
\\
\multicolumn{14}{c}{The 8 EISERS candidates with only a lower limit to $\alpha$(150-325 MHz)}\\
\\
J0045+8810	&		&		&		&		&		&		&		&	15.157	&	15.006	&	$>$12.541	&	$>$9.447	&		&		\\
J0304+7727	&	QSO	&		&	23.98	&	23.63	&	21.75	&	21.19	&	20.19	&	16.013	&	15.635	&	12.982	&	$>$9.379	&	0.35	&	2.23	\\
J1326+5712	&	QSO	&		&	23.11	&	21.79	&	20.61	&	19.48	&	19.12	&	14.607	&	13.872	&	11.408	&	8.503	&	1.32	&	2.5	\\
J1536+8154	&	QSO	&		&		&		&		&		&		&	17.363	&	16.168	&	$>$13.505	&	$>$9.679	&		&		\\
J1658+4732	&		&		&	23.91	&	23.41	&	22.16	&	21.61	&	20.99	&	17.49	&	16.78	&	$>$13.21	&	$>$9.356	&	0.5	&	1.75	\\
J1722+7046	&	QSO	&	1.955	&	19.67	&	19.5	&	19.39	&	19.26	&	19.18	&	16.485	&	15.096	&	12.028	&	9.468	&	0.17	&	0.28	\\
J1723+7653	&	QSO	&	0.680	&	18.75	&	18.34	&	18.22	&	18.04	&	17.89	&	13.222	&	12.213	&	9.329	&	6.976	&	0.41	&	0.53	\\
J2317+4738	&		&		&		&		&		&		&		&	15.652	&	15.825	&	$>$12.467	&	$>$9.368	&		&		\\

\hline
\end{tabular}

{$^{\dagger}$\citet{Souchay2015}}\\
{$^{\ddagger}$ SDSS -- http://skyserver.sdss.org/dr14/ ; WISE --\citet{WISE2013}.}\\
\end{table}
\clearpage

\newpage
\clearpage

\section{Results and Discussion}

From Figs~\ref{fig:spec_all_detected} \& ~\ref{fig:spec_all_undetected}, at least 8 of the total 15 sources in the present list of EISERS candidates show a spectrum of GPS type\footnote{GPS sources are defined as having an integrated spectrum that shows a single peak near 1 GHz and steep slopes on either side of the peak, cf. \citep{Spoelstra1985, GopalKrishna1983, GopalKrishna1993, deVries1997, Odea1998, An2012}}. Nearly always, the spectral peak for these sources lies within the 0.3 to 1.5 GHz range, consistent with the canonical GPS spectrum \citep{deVries1997}. For only one source, J0304+7727, the available flux density measurements might not encompass the turnover frequency, though the data in the plot can still be reconciled with the spectral peak occurring near 1 GHz (Fig.~\ref{fig:spec_all_undetected}). For a few of the sources, namely, J0847+5723, J1326+5712 and J1430+3649, and, possibly, J1549+5038, the available radio spectrum provides evidence for a second, more compact component which becomes transparent (and often, even dominant) above a few GHz (Figs~\ref{fig:spec_all_detected} \& ~\ref{fig:spec_all_undetected}; see below).

\subsection*{\textbf{Comments on Individual Sources}}
\subsection*{\textbf{J0304+7727}}
The VLBA image at 8.6 GHz shows this source to consist of a pair of compact components separated by $\sim$10 mas \citep{Sokolovsky2011}. The 3 available spectral measurements fall short of pinning down the turnover frequency, but it is unlikely to be below 0.5 GHz (Fig.~\ref{fig:spec_all_undetected}).

\subsection*{\textbf{J0754+5324}}
The classical GPS shape of its radio spectrum (Fig~\ref{fig:spec_all_detected}) hints at a compact symmetric object (CSO) type morphology \citep{Phillips1982, Conway1994, Wilkinson1994}. Indeed, this appears consistent with the 5 GHz VLBA image which shows an overall size of $\sim$9 mas \citep{Helmboldt2007}. An earlier VLBI image at 8.4 GHz showed the source to be dominated by two hot spots with a strong jet pointing toward the northernmost lobe \citep{Peck2000}. The core appears to be a knot in the jet, speeding toward the northwestern hot spot at 0.060${\pm}$0.026 mas per year \citep{Gugliucci2005}.

\subsection*{\textbf{J0847+5723}}
The VLBI image at 5 GHz shows a triple source, consistent with a CSO morphology\citep{Xu1995}. Such an interpretation would be in accord with its GPS type spectrum (Fig.~\ref{fig:spec_all_detected}). The source is also detected at 15 GHz (66.9${\pm}$6.7 mJy \citet{Taylor2005}) and at 30 GHz (43${\pm}$04 mJy \citet{Lowe2007}), suggesting a compact component becoming transparent at these frequencies (Fig.~\ref{fig:spec_all_detected}).

\subsection*{\textbf{J0858+7501}}
This source was published as part of the List 3 of GPS sources \citep{GopalKrishna1993}.

\subsection*{\textbf{J1326+5712}}
The 5 GHz VLBA image of this source made under the VIP survey \citep{Helmboldt2007} shows a complex morphology with an overall size of 4.2 mas. The source was found to be unpolarised in the 10.5 GHz observations with the Effelsberg telescope \citep{Pasetto2016}. The spectrum shows an upturn around 8 GHz (Fig.~\ref{fig:spec_all_undetected}), which implies the existence of an even more compact component whose spectrum peaks at frequency $>$20GHz \citep{Snellen1995}. 

\subsection*{\textbf{J1430+3649}}
The 5 GHz VLBA image of this source made under the VIP survey \citep{Helmboldt2007} shows it to be single source, with a mean radius of 1.3 mas. However, this deceptively simple radio morphology masks the complexity of its radio spectrum which shows a second even more component which dominates above 2-3 GHz and remains opaque at least upto 15 GHz (Fig.~\ref{fig:spec_all_detected}). 

\subsection*{\textbf{J1549+5038}}
The VLBA images at 5 GHz \citep{Xu1995,Helmboldt2007} and at 2 \& 8 GHz \citep{Fey2000} have resolved this source into a dominant flat-spectrum core and a fairly bright curved jet $\sim$10 mas long and extending to the south-west. The jet is resolved into 2-3 knots. The source is also detected at 15 GHz (810${\pm}$03 mJy, \citet{Richards2011}) and at 30 GHz (738${\pm}$39 mJy, \citet{Lowe2007}) which is the highest frequency up to the spectrum of any source in our sample is seen to remain opaque (Figs~\ref{fig:spec_all_detected} \& ~\ref{fig:spec_all_undetected}). This is also in accord with the observed strong variability at 15 GHz  (modulation index  9\%, see \citet{Richards2014}). \citet{Pasetto2016} have inferred a high (rest-frame) rotation measure (RM = 1400$\pm$500 rad $m^{-2}$) for this source, revealing the presence of dense magneto-ionic plasma.

\subsection*{\textbf{J1700+3830}}
The 5 GHz VLBA image of this source, made under the VIP survey \citep{Helmboldt2007}, shows it to be a core with a jet towards the east, which possibly terminates in a lobe $\sim$50 mas away from the core. It has a classical GPS spectrum which peaks near 0.5 GHz (Fig.~\ref{fig:spec_all_detected}).

\subsection*{}    
As we have discussed in Paper I and Paper II, an inverted radio spectrum with a slope $\alpha$ $>$ $\alpha_c$ $=$ +2.5 would be inconsistent with the standard SSA interpretation of the spectral turnover, if applied to the canonical (i.e., power-law) energy distribution of the relativistic electrons in the source \citep{Slish1963, Kellermann1969,Pacholczyk1970, Rybicki1979}. Instead, the SSA scenario would require a non-standard energy distribution for the relativistic electrons, e.g., a delta-function or a Maxwellian \citep{Rees1967}, or, alternatively, a large excess of electrons at lower energies, over a power-law extrapolated from higher energies \citep{deKool1989}. An alternative, less radical, possibility to explain the ultra-steep spectral turnover is to invoke free-free absorption (FFA) \citep{Kellermann1966, Bicknell1997, Kuncic1998, Kameno2000, Vermeulen2003, Stawarz2008, Callingham2015}. In fact, from VLBI studies, a substantial evidence for FFA effects in extragalactic radio sources is already available (albeit, only for the nuclear region). Prominent examples of flux attenuation on parsec scale due to FFA include 3C 345 \citep{Matveenko1990}, Centaurus A \citep{Jones1996,Tingay2001}, Cygnus A \citep{Krichbaum1998} and 3C 84 \citep{Walker1994, Levinson1995}. Whether FFA can attenuate most of the radio emission of a source, as inferred from the sharp turnover of the integrated radio spectrum (Figs~\ref{fig:spec_all_detected} \& ~\ref{fig:spec_all_undetected}), remains to be demonstrated. In this context it is interesting that at least for one source (J1549+5038) in the present list of EISERS candidates, there is evidence for an unusually large rotation measure (RM), signalling the presence of dense thermal plasma. Interestingly, the radio spectrum of this source shows a very prominent, presumably younger, second component emerging above $\sim$5 GHz (Fig.~\ref{fig:spec_all_detected}), which has been identified as a blazar (see above).
 
Finally, it has been pointed out by several authors that radio flux densities of GPS sources show minimal variability, particularly at low frequencies \citep{Rudnick1982, Seielstad1983, Fassnacht2001, Callingham2015}. This is nicely corroborated by the internal consistency found among the multiple (non-simultaneous) flux density measurements at close-by frequencies, that are available for some of the 15 sources in our sample (see Figs~\ref{fig:spec_all_detected} \& ~\ref{fig:spec_all_undetected}). We note that J1700+3830 is the only source where a significant flux variation (near the spectral peak) may have occurred. It is widely held that a prime cause of radio flux variation at metre wavelengths is refractive scintillations due to plasma irregularities in the interstellar medium of our galaxy \citep{Rickett1990, McGilchrist1990} and references therein, also \citet{Bell2014}. The resulting, flux variability at metre wavelengths, is expected to be broad-band and hence probably a minor factor in the spectral index determination over the (narrow) frequency range encompassing the observed most inverted spectral region \citep{Rickett1990}. As mentioned above, the non-variability is particularly unexpected for the sources displaying a GPS spectrum. Indeed, it is perhaps not a mere coincidence that the two best EISERS candidates out of the seven presented in our Paper I (in which the radio spectra were drawn using non-simultaneous flux measurements) are also the ones that came out on the top when quasi-simultaneous multi-frequency flux measurements of those 7 candidates were subsequently made with the GMRT (Paper II). Hence, we expect that for many of the present EISERS candidates, neither the shape of the low-frequency spectral turnover (see, Figs~\ref{fig:spec_all_detected} \& ~\ref{fig:spec_all_undetected}), nor the estimates of $\alpha$ (150-325 MHz) (Table~\ref{table:spec-prop}) is likely to be found much in error on account of the non-simultaneity of the multi-frequency flux measurements. Thus, even though, determination of the radio spectra via simultaneous flux density measurements at multiple frequencies, clearly remains a cherished goal, we are reasonably confident that, when available, such contemporaneous data will validate a majority of the EISERS candidates we have listed in Table~\ref{table:spec-prop}. A second priority would be to complete the missing redshift and rotation measure information for this sample, particularly in view of its importance in assessing the role of FFA in the EISERS phenomenon.

\section*{Acknowledgements}
The Giant Metrewave Radio Telescope (GMRT) is a national facility operated by the National Centre for Radio Astrophysics (NCRA) of the Tata Institute of Fundamental Research (TIFR).  We thank the staff at NCRA and GMRT for their support. This research has used the TIFR. GMRT. Sky. Survey (http://tgss.ncra.tifr.res.in) data products, NASA's Astrophysics Data System and NASA/IPAC Extragalactic Database (NED), Jet Propulsion Laboratory, California Institute of Technology under contract with National Aeronautics and Space Administration and VizieR catalogue access tool, CDS, Strasbourg, France. We thank the NRAO staff for providing AIPS. SP would like to thank DST INSPIRE Faculty Scheme (grant code: IF-12/PH-44) for funding his research group

\newpage 
\bibliography{references}  
\end{document}